\documentclass{amsart}
\usepackage{amsbsy,amssymb,amscd,amsfonts,latexsym,amstext,delarray,
amsmath,amsthm,graphicx}

\input xypic
\usepackage[all]{xy}

\newtheorem{thm}{Theorem}[section]
\newtheorem{prop}[thm]{Proposition}
\newtheorem{cor}[thm]{Corollary}
\newtheorem{lem}[thm]{Lemma}

\numberwithin{equation}{section}

\def\bT{{\mathbb T}}

\def\cA{{\mathcal A}}
\def\cB{{\mathcal B}}

\def\cE{{\mathcal E}}
\def\cF{{\mathcal F}}

\def\cH{{\mathcal H}}

\def\cL{{\mathcal L}}

\def\cO{{\mathcal O}}

\def\cS{{\mathcal S}}
\def\cT{{\mathcal T}}

\def\cZ{{\mathcal Z}}

\def\C{{\mathbb C}}
\def\F{{\mathbb F}}

\def\Q{{\mathbb Q}}
\def\Z{{\mathbb Z}}
\def\R{{\mathbb R}}

\def\fA{{\mathfrak A}}

\def\Aut{{\rm Aut}}

\def\End{{\rm End}}

\def\SL{{\rm SL}}

\def\Tr{{\rm Tr}}

\title{Codes as fractals and noncommutative spaces}
\author{Matilde Marcolli}
\author{Christopher Perez}
\address{Department of Mathematics \\ California Institute of Technology \\  Pasadena \\ CA 91125 \\ USA}
\email{matilde@caltech.edu}
\email{cperez@caltech.edu}

\begin{document}
\maketitle

\begin{abstract}
We consider the CSS algorithm relating self-orthogonal classical linear codes
to q-ary quantum stabilizer codes and we show that to such a pair of a classical and
a quantum code one can associate geometric spaces constructed using 
methods from noncommutative geometry, arising from rational noncommutative tori
and finite abelian group actions on Cuntz algebras and fractals associated 
to the classical codes.
\end{abstract}

\section{Introduction}

This paper contains a series of simple observations on the theme of error-correcting
codes, looked at from the point of view of fractal and noncommutative geometry
advocated recently in \cite{ManMar}. In particular, in this paper we extend that point
of view to include quantum stabilizer codes and their relation to classical linear 
codes through the CSS algorithms. 

\smallskip

After reviewing some basic facts about the CSS algorithm relating 
self-orthogonal classical linear codes to q-ary quantum stabilizer codes
in this section, we show in \S \ref{QtoriSec} that the construction of
q-ary quantum stabilizer codes can be naturally expressed in terms
of the geometry of rational noncommutative tori. We also show that,
if the q-ary quantum stabilizer codes is obtained from a classical
self-orthogonal linear code via the CSS algorithm, then some properties
of the classical code can be seen in the resulting algebra, such as a
filtration that corresponds to the Hamming weight. 

\smallskip

In \S \ref{AlgCQcodeSec}, we recall some results of \cite{ManMar} on
associating  to a classical code $C$ a fractal $\Lambda_C$ and an
operator algebra, a Cuntz algebra $\cO_C$ or a Toeplitz algebra $\cT_C$.
We give an explicit example of a very simple code for which one can
completely visualize the associated fractal space. Since we are dealing
only with linear codes here, unlike in the more general setting of
\cite{ManMar}, we can enrich these spaces and algebras with
group actions coming from the linear structure of the code. We show
that one obtains in this way a crossed product algebra that has the
Rokhlin property. We comment on other possible related actions on
can consider on the fractal $\Lambda_C$, such as adding machines.
We then show that the fractals of classical codes can be embedded,
compatibly with the group actions in a disconnection of a torus and
that the geometric construction via rational noncommutative
tori obtained in the previous section can be pulled back to the
fractal $\Lambda_C$ via this embedding and the projection 
from the disconnection to the torus giving rise to a quotient
space by the group action which is a fibration over a torus with 
fiber a fractal.  We also show how one can use a crossed product
algebra defined by the action of $(\Z/p\Z)^2$ on the disconnection
of the torus $T^2$ to obtain a noncommutative space with the
property that all the noncommutative spaces associated to
individual classical codes via the group action on the associated
fractal $\Lambda_C$ can be embedded inside (powers of) this
universal one. This gives a common space inside which to compare
noncommutative spaces of different codes and relate their properties.
We hope this may be useful in optimization problems. 
We also give a reinterpretation of the weight polynomial of a 
linear code in terms of subfractals of $\Lambda_C$ and 
multiplicities of embeddings of the corresponding Toeplitz
algebras. 

\smallskip

We conclude in \S \ref{RemkSec} with some brief remarks on 
methods and recent results in noncommutative geometry that 
may be applied to the study of the correspondence between
classical and quantum codes via the geometric spaces
and algebras we describe in this paper.

\subsection{Classical linear codes}

We recall briefly the general setting of classical codes, following \cite{TsfaVla}.
An alphabet is a finite set $\fA$ of cardinality $q\geq 2$. A classical 
code is a subset $C\subset \fA^n$.
Elements of $C$ are code words, identified with n-tuples $x=(a_1,\ldots,a_n)$
in $\fA^n$. 

\smallskip

We set $k =k(C) = \log_q \# C$ and $\lfloor k \rfloor$ the integer part of $k$.
The code rate or {\em transmission rate} of the code is the ratio $R =k/n$.

\smallskip

The Hamming distance between two code words $x=(a_i)$ and $y=(b_i)$
is given by $d(x,y)=\# \{ i \,|\, a_i\neq b_i \}$. The {\em minimum distance}
$d=d(C)$ of the code is given by $d(C)=\min \{ d(x,y) \,|\, x,y\in C, \, x\neq y \}$.
The {\em relative minimum distance} of the code is the ratio $\delta = d/n$.

\smallskip

A classical code $C$ with these parameters is called an $[n,k,d]_q$ code.

\smallskip

The most important class of codes, in the classical setting, is given by the
{\em linear codes}. In this class, the alphabet is given by the elements
of a finite field $\fA=\F_q$ of cardinality $q=p^r$ and characteristic $p>0$. 
The code is linear if $C\subset \F_q^n$ is an $\F_q$-linear subspace of
the vector space $\F_q^n$. In particular $k=\lfloor k \rfloor$ is an integer for
linear codes and is the dimension of $C$ as a vector space.

\smallskip

Given an $\F_q$-bilinear form $\langle\cdot,\cdot\rangle$ on $\F_q^n$,
a code $C\subset \F_q^n$ is {\em self-orthogonal} if, for all code words
$x,y\in C$ one has $\langle x, y \rangle =0$.
The dual code $C^\perp$ is given by the set of vectors $v$ in $\F_q^n$ 
satisfying $\langle v, x \rangle =0$ for all $x\in C$. Thus, a self-orthogonal
code satisfies $C\subseteq C^\perp$.

\subsection{Quantum stabilizer codes}

A qbit is a vector in the finite dimensional Hilbert space $\C^2$. Quantum
codes as in \cite{Shor} have been typically constructed over qbit spaces 
$(\C^2)^{\otimes n}$. These are referred to as binary quantum codes. 
However, more recently nonbinary quantum codes have also been 
constructed \cite{Rains}, \cite{CRSS}, 
especially in relation to classical codes associated to algebraic curves.

In this more general setting of nonbinary quantum codes, one considers
a vector $\C^q$ representing the states of a q-ary system.  A {\em q-ary quantum
code} of {\em length} $n$ and {\em size} $k$ is then a $k$-dimensional
$\C$-linear subspace of $\C^{q^n}=(\C^q)^{\otimes n}$. A quantum error
is a linear map $E \in \End_\C(\C^{q^n})$. For a quantum error of the form
$E = E_1 \otimes \cdots \otimes E_n$, the weight is $w(E)=\# \{ i \,|\, E_i \neq id \}$. 
A quantum error $E$ is {\em detectable} by a quantum code $Q$ if
$P_Q E P_Q =\lambda_E P_Q$, where $P_Q$ is the orthogonal projection onto $Q
\subset \C^{q^n}$ and $\lambda_E \in \C$ is a constant depending only on $E$.
The {\em minimum distance} of a quantum code $Q$ is 
\begin{equation}\label{dQ}
 d_Q = \max \{ d \, | \, E \text{ is detectable } \,\, \forall  E=E_1\otimes \cdots \otimes E_n \text { with } w(E)\leq d-1 \} .
\end{equation} 
A quantum codes with these parameters is called a $[[ n, k, d ]]_q$ quantum code.

\smallskip

We recall the following notation and basic facts following \cite{AshKnill}. Let $q=p^m$ and
consider, as above, the field $\F_q$. Viewed as an $\F_p$-vector space,
it can be identified, after choosing a basis, with $\F_p^m$. Thus, given an element 
$x\in \F_q^n$, $x=(a_1,\ldots,a_n)$, we can identify the coefficients 
$a_i\in \F_q$ with vectors $a_i =(a_{i1},\ldots, a_{im})$ with $a_{ij}$ 
in $\F_p$. These in turn can then be thought of as elements of $\Z/p\Z$,
that is, as integer numbers $0\leq a_{ij} \leq p-1$. Thus, given a
linear operator $L \in \End_\C(\C^p)$, such that $L^p=id$,
we can consider the integer powers $L^{a_{ij}}$. 

\smallskip

In particular, consider the two operators $T$ and $R$ on $\C^p$ given by the matrices
\begin{equation}\label{Top}
T =\left( \begin{array}{cccccc}
0 & 1 & 0 & \cdots & 0 & 0 \\
0 & 0 & 1 & \cdots & 0 & 0 \\
\vdots & & & & & \vdots \\
0 & 0 & 0 & \cdots & 0 & 1 \\
1 & 0 & 0 & \cdots & 0 & 0 
\end{array}\right)
\end{equation}
\begin{equation}\label{Rop}
R = \left( \begin{array}{cccccc}
1 & 0 & 0 & \cdots & 0 & 0 \\
0 & \xi & 0 & \cdots & 0 & 0 \\
0 & 0 & \xi^2 & \cdots & 0 & 0 \\
\vdots & & & & & \vdots \\
0 & 0 & 0 & \cdots & \xi^{p-2} & 0 \\
0 & 0 & 0 & \cdots & 0 & \xi^{p-1}
\end{array}\right),
\end{equation}
where $\xi = \exp(2\pi i /p)$. These have the properties that
\begin{equation}\label{TRrel1}
T^p = R^p = id \ \ \ \ \text{ and } \ \ \ \  TR =\xi RT,
\end{equation}
which also imply the relations
\begin{equation}\label{TRrel2}
T^k R^\ell = \xi^{k\ell} R^\ell T^k \ \ \  \text{ and } \ \ \ 
(T^k R^\ell) (T^r R^s) = \xi^{-r\ell} T^{r+k} R^{s+\ell} =
\xi^{sk-r\ell}  (T^r R^s)  (T^k R^\ell) .
\end{equation}
Moreover, the operators $T^k R^\ell$ form an orthonormal basis of $M_p(\C)=\End_\C(\C^p)$
with respect to the inner product $\langle A, B\rangle =\Tr(A^* B)$.

\smallskip

Consider then linear maps $E=E_1\otimes\cdots\otimes E_n$ in $\End_\C(\C^{q^n})$,
with $q=p^m$, where the factors $E_i$ are of the form $E_i = T_x R_y$, where 
$x$ and $y$ are elements  in $\F_q$,
which we write as vectors $x=(a_1,\ldots,a_m)$, $y=(b_1,\ldots, b_m)$ with coefficients
$a_i$ and $b_i$ in $\F_p$, and we set 
$T_x= T^{a_1}\otimes \cdots \otimes T^{a_n}$ and $R_y=R^{b_1} \otimes\cdots
\otimes R^{b_n}$, with the same conventions explained above and with $T$ and $R$
as in \eqref{Top} and \eqref{Rop}. Thus, for $v=(x_1,\ldots, x_n)$ and $w=(y_1,\ldots, y_n)$
vectors in $\F_q^n$, we can write a corresponding operator
\begin{equation}\label{Evw}
E_{v,w} = T_{x_1}R_{y_1}\otimes \cdots \otimes T_{x_n} R_{y_n}.
\end{equation}
The relations \eqref{TRrel1} and \eqref{TRrel2} imply that
\begin{equation}\label{TRrel3}
E_{v,w} E_{v',w'} = \xi^{\langle v,w'\rangle - \langle w,v' \rangle} E_{v',w'} E_{v,w},
\end{equation}
where, for $v,w \in \F_q^n$, the bilinear form $\langle v,w\rangle$ is defined as 
\begin{equation}\label{bilin}
 \langle v,w\rangle = \sum_{i=1}^n \sum_{j=1}^m a_{ij} b_{ij}. 
\end{equation} 
Similarly, one also has
\begin{equation}\label{TRrel4}
E_{v,w} E_{v',w'} = \xi^{-\langle w,v' \rangle} E_{v+v', w+w'},
\end{equation}
and $E_{v,w}^p =id$ as a $p^{nm}\times p^{nm}$ matrix.

\smallskip

One then denotes by $\cE$ (see \cite{AshKnill}) the subgroup of $\Aut_\C(\C^{q^n})$
given by the invertible linear maps of the form
\begin{equation}\label{Egroup}
\cE =\{ \xi^k E_{v,w} \, |\, v,w\in \F_q^n, \, 0\leq k \leq p-1 \}.
\end{equation}
It is a finite group of order $p^{2mn+1}$. The center $\cZ$ of $\cE$ is the subgroup
$\{ \xi^k \, id \}$ isomorphic to $\Z/p\Z$.

\smallskip

A {\em quantum stabilizer code}  is a quantum code that
is obtained as joint eigenspace of all the linear transformations
in a commutative subgroup of $\cE$. Namely, let $\cS \subset \cE$
be a commutative subgroup with $\# \cS = p^{r+1}$,
and let $\chi: \cS \to U(1)$ be a
character that is trivial on $\cZ$. Then the associated quantum stabilizer
code $Q=Q_{\cS,\chi}$ is given by the linear subspace of $\C^{q^n}$
\begin{equation}\label{QSchi}
Q_{\cS,\chi} = \{ \psi\in  \C^{q^n} \,|\,  A \psi = \chi(A) \psi, \,\, \forall A\in \cS \}.
\end{equation}
The dimension of this vector space is $p^{mn-r}$, see \cite{AshKnill}.

\subsection{Classical and quantum codes}
A very interesting aspect of quantum stabilizer codes is that there is an
efficient procedure to go back and forth between classical self-orthogonal
linear codes and quantum stabilizer codes with a good control
over the respective parameters. The procedure is explained in detail in
\cite{AshKnill} and we only recall it here briefly for what we will need 
to use later in the paper.

\smallskip

Given a quantum stabilizer code $Q=Q_{\cS,\chi}$ as above and an 
$\F_p$-linear automorphism $\varphi \in \Aut_{\F_p}(\F_p^m)$, the
set
\begin{equation}\label{CofQ}
C=C_{Q,\varphi} = \{ (v,\varphi^{-1}(w))\,|\, E_{v,w}\in \cS \}
\end{equation}
is an $\F_p$-linear code of length $2n$, with $\# C =p^r$, where
$\# \cS =p^{r+1}$. It is self-orthogonal with respect to the bilinear form
$\langle v,\varphi(w')\rangle - \langle v',\varphi(w) \rangle$,
with $\langle v,w\rangle$ as in \eqref{bilin}. The minimum distance $d_Q$ of
the quantum stabilizer code $Q_{\cS,\chi}$ is related to the classical code 
by  $d_Q = d^\perp=d_{C^\perp \smallsetminus C} :=\min \#\{ i \,|\, v_i \neq 0 \, \text{ or } \,
w_i \neq 0 , \,\,  (v,w) \in \F_q^{2n}, \, (v,w) \in C^\perp\smallsetminus C \}$.

\smallskip

Conversely, given a classical linear self-orthogonal code in $\F_q^{2n}$,
with $\# C=p^r$, the linear maps $E_{v,\varphi(w)}$, with $(v,w)$ ranging over 
an $\F_p$-basis of $C$, together with the elements $\xi^k id$, generate 
a subgroup $\cS$ of $\cE$. The self-orthogonal condition implies by \eqref{TRrel4}
that the subgroup $\cS$ is abelian. By construction, it is of order $\# \cS =p^{r+1}$.
The associated quantum stabilizer codes $Q_{\cS,\chi}$ then have
parameters $[[n,n-r/m,d^\perp]]_q$.

\smallskip

Notice how, in this construction, the field extension $\F_q$ of $\F_p$ is identified with 
the vector space $\F_p^m$, without keeping track of the field structure. The only choice
in the data that can be arranged so as to remember the remaining structure is the
automorphism $\varphi$. Namely, as shown in \cite{AshKnill}, that can be chosen 
so that the bilinear form becomes 
$\Tr(\langle v,w'\rangle - \langle v',w \rangle)$ with $\langle v,w \rangle=\sum_{i=1}^n
v_i w_i$, with the product in the field $\F_q$ and $\Tr: \F_{p^m} \to \F_p$ the standard trace
$\Tr(x)=\sum_{k=0}^{m-1} x^{p^k}$.

\smallskip

This procedure that constructs quantum stabilizer codes from classical
self-orthogonal linear codes was further refined in \cite{KimWal}, but for
our purposes here this description suffices. 

\section{Quantum codes and rational noncommutative tori}\label{QtoriSec}

In this section we show that the data of quantum stabilizer codes described
above can also be described in terms of rational noncommutative tori.

\subsection{Twisted group rings}

We recall here also something about twisted group rings, which will be
useful later. Given a discrete group $G$, the group ring $\C[G]$ admits
a (reduced) $C^*$-completion $C^*_r(G)$ by taking the closure of $\C[G]$ 
in the operator norm of the algebra of bounded operators $\cB(\ell^2(G))$, 
for the action of $\C[G]$ on $\ell^2(G)$ by $r_g f(g') = f(g'g)$. 
A multiplier  $\sigma: G \times G \to U(1)$ is a 2-cocycle satisfying
the conditions $\sigma(g,1)=\sigma(1,g)=1$ and
$\sigma(g_1,g_2)\sigma(g_1g_2,g_3)=
\sigma(g_1,g_2g_3) \sigma(g_2,g_3)$.
The twisted group ring $\C[G,\sigma]$ is generated by the twisted translations
$r_g^\sigma f(g')=f(g'g)\,\sigma(g',g)$. The properties of the multiplier
ensure that the resulting algebra is still associative. The composition
of twisted translations is given by $r^\sigma_g r^\sigma_{g'} = \sigma(g,g') r^\sigma_{gg'}$.
The twisted (reduced) group $C^*$-algebra $C^*_r(G,\sigma)$ is the norm 
closure of $\C[G,\sigma]$ in $\cB(\ell^2(G))$.

\smallskip

The following simple observation relates these general facts to the codes we recalled
in the previous section.

\begin{lem}\label{MCpsigma}
For $q=p^m$, the matrix algebra $M_{q^n}(\C)$ can be identified 
with the twisted group $C^*$-algebra $C^*((\Z/p\Z)^{2mn},\sigma)$, where the 
multiplier $\sigma: (\Z/p\Z)^{2m}\times (\Z/p\Z)^{2m} \to U(1)$ is given by 
\begin{equation}\label{Matsigma}
 \sigma( (v,w), (v',w') ) = \xi^{-\langle w, v' \rangle}, 
\end{equation}
with $\langle\cdot,\cdot\rangle$ defined as in \eqref{bilin} and with $\xi=\exp(2\pi i/p)$.
This is, in turn, the $C^*$-algebra $C^*(\cE)$, with $\cE$ as in \eqref{Egroup}, generated by the transformations $E_{v,w}$ of \eqref{Evw}.
\end{lem}

\proof The expression \eqref{Matsigma} defines a multiplier on $(\Z/p\Z)^{2mn}$. In fact,
$\sigma((v,w),(0,0))=\sigma((0,0),(v,w))=1$ and $$\sigma((v,w),(v',w'))
\sigma((v+v',w+w'),(v'',w'')) = \xi^{-\langle w,v'\rangle-\langle w,v'' \rangle - \langle w',v'' \rangle} $$
$$ = \sigma((v,w),(v'+v'',w'+w''))\sigma((v',w'),(v'',w'')).$$
The twisted group $C^*$-algebra (which is the same as the twisted group ring in
this finite dimensional case) $C^*((\Z/p\Z)^{2mn},\sigma)$ then has generators 
$r_{(v,w)}^\sigma$ such that $r_{(v,w)}^\sigma r_{(v',w')}^\sigma =
\xi^{-\langle w, v' \rangle} r_{(v+v',w+w')}^\sigma$. By direct comparison with
\eqref{TRrel4}, one sees that the identification $r_{(v,w)}^\sigma \mapsto E_{v,w}$
identifies $C^*((\Z/p\Z)^{2mn},\sigma)$ with $C^*(\cE/\cZ)$. In fact, notice that the
relation \eqref{TRrel3} also follows from the twisted group ring relations since we
obtain 
$$ r^\sigma_{(v,w)} r^\sigma_{(v',w')} = \sigma((v,w),(v',w')) \sigma((v',w'),(v,w))^{-1} 
r^\sigma_{(v',w')} r^\sigma_{(v,w)} $$
which then gives relation \eqref{TRrel3}. The identification between $C^*(\cE/\cZ)$
and $M_{q^n}(\C)$ follows from the known fact that the transformations $E_{v,w}$
generate $\End_\C((\C^q)^{\otimes n})$.
\endproof

\subsection{Rational noncommutative tori}

The (rational or irrational) rotation algebras, also known as noncommutative tori, 
are the most widely studied examples of noncommutative spaces. As a $C^*$-algebra,
the rotation algebra $\cA_\theta$ is generated by two unitaries $U$ and $V$, subject 
to the commutation relation
\begin{equation}\label{UVcomm}
UV = \xi VU,
\end{equation}
with $\xi = \exp(2 \pi i \theta)$. In the rational case, $\theta \in \Q$, it is well
known that these algebras are Morita equivalent to the commutative algebra of
functions $C(\bT^2)$ on the ordinary commutative torus $\bT^2$, while in the irrational
case $\theta\in \R\smallsetminus \Q$, the Morita equivalence classes correspond
to the orbits of the action of $\SL_2(\Z)$ on the real line by fractional linear transformations.

\smallskip

Let us look more closely at the rational case with $\xi = \exp(2\pi i /p)$. Then
elements in the rotation algebra $\cA_{1/p}$ are of the form
\begin{equation}\label{eltsAp}
\cA_{1/p} \ni a = \sum_{k,\ell} f_{k,\ell}(\mu,\lambda) \, T^k R^\ell,
\end{equation}
where $f_{k,\ell}(\mu,\lambda)$ are continuous functions of $(\lambda,\mu)\in S^1\times S^1=\bT^2$
and $T$ and $R$ are the matrices \eqref{Top} and \eqref{Rop}. The sum is a finite sum for
$0\leq k,\ell \leq p-1$ since $T^p=R^p=id$. In particular, the generators $U$ and $V$ are
given, respectively, by $U=\mu T$ and $V=\lambda R$, with $\mu=\exp(2\pi i t)$ and
$\lambda =\exp(2\pi i s)$ in $S^1$. 
To see this notice that the algebra $\cA_{1/p}$ is generated by elements of the
form $$\sum_{k,\ell \in \Z} a_{k\ell} U^k V^\ell.$$ Since $T^p=R^p=id$, we can rewrite
these as $$\sum_{k,\ell \in \Z/p\Z} \sum_{k',\ell' \in \Z} a_{k+k'p, \ell + \ell' p}
\mu^{k+k'p} \lambda^{\ell + \ell'p}  T^k R^\ell =\sum_{k,\ell \in \Z/p\Z} f_{k,\ell}(\lambda,\mu)
T^k R^\ell. $$

\smallskip

\subsection{Quantum codes and vector bundles}

Recall (see \cite{GBVF}, Proposition 12.2) that the rational noncommutative torus
$\cA_{n/m}$ is isomorphic to the algebra $\Gamma(T^2,\End(E_m))$ of sections 
of the endomorphism bundle of a rank $m$ vector bundle $E_m$ over the ordinary
torus $T^2$, obtained as follows. Consider the trivial bundle over $T^2$ with
fiber $M_m(\C)$, with the action of $(\Z/m\Z)^2$ given by
$$ \tau_{1,0}: (\mu,\lambda, M) \mapsto (\mu, e^{-2\pi i n/m}\lambda, T M T^{-1}), \ \ \ 
\tau_{0,1}:  (\mu,\lambda, M) \mapsto (e^{2\pi i n/m}\mu, \lambda, R M R^{-1}) . $$
The quotient by this action defines a non-trivial bundle over $T^2$, which we can
view as the endomorphism bundle $\End(E_m)$ of a vector bundle $E_m$ of rank $m$,
with fiber $M_m(\C)$. The algebra of sections $\Gamma(T^2,\End(E_m))$ is by
construction the fixed point subalgebra of the algebra $C(T^2,M_m(\C))=C(T^2)\otimes 
M_m(\C)$ of endomorphisms of the trivial bundle, under the action of $(\Z/m\Z)^2$ 
described above. The above action gives on the algebra $C(T^2)\otimes 
M_m(\C)$ the action 
\begin{equation}\label{actalpha}
\begin{array}{c}
 \alpha_{1,0}:  
f(\mu,\lambda) \otimes M \mapsto f(\mu, e^{-2\pi i n/m}\lambda) \otimes T M T^{-1}, \\[2mm]  
\alpha_{0,1}:  f(\mu,\lambda) \otimes M \mapsto f(e^{2\pi i n/m}\mu, \lambda) 
\otimes R M R^{-1} . 
\end{array}
\end{equation}
The fixed point subalgebra is then generated by the elements $\mu \otimes T$ 
and $\lambda \otimes R$, which satisfy the commutation relation of the 
generators $U$ and $V$ of the noncommutative torus, and is therefore
isomorphic to $\cA_{n/m}$. In particular, there is a $C^*$-algebra homomorphism
$\cA_{n/m} \to M_m(\C)$ that sends the generators $U$ and $V$ to the matrices
$T$ and $R$.

\smallskip

We then use this description of the rational noncommutative tori to give
a geometric interpretation of the data of quantum stabilizer codes.

\begin{prop}\label{QcodesNCtori}
Let $E_p$ be the rank $p$ bundle over $T^2$ such that $\cA_{1/p}=\Gamma(T^2,\End(E_p))$.
Then, for $q=p^m$, a $q$-ary quantum stabilizer code $Q_{\cS,\chi}$ of length $n$ and size $k$
corresponds to a subalgebra $\cA_{\cS}\subset \cA_{1/p}^{\otimes r}$, with $r=nm$, 
and subbundle $\cF_{\cS,\chi}$ of the external tensor product $E_p^{\boxtimes mn}$ over
$T^{2r}$, on which the elements of the algebra $\cA_\cS$ act as scalars. 
Conversely, these data determine
a $q$-ary quantum stabilizer code $Q_{\cS,\chi}$ of length $n$ and size $k$.
\end{prop}

\proof Let us first consider the tensor product algebra $C(T^2, M_p(\C))^{\otimes r}$
where $r=mn$. We can write this also as $(C(T^2)\otimes M_p(\C))^{\otimes r}=C(T^{2r})
\otimes M_{q^n}(\C)=C(T^{2r},M_{q^n}(\C))$, for $q=p^m$. This is therefore the
algebra of endomorphisms of the trivial bundle with fiber $\C^{q^n}$ over the higher
dimensional torus $T^{2r}$. The action of $(\Z/p\Z)^2$ on $C(T^2, M_p(\C))$ given in
\eqref{actalpha} extends to an action of $(\Z/p\Z)^{2r}$ on $C(T^{2r},M_{q^n}(\C))$, which
is given by 
\begin{equation}\label{actalphavw}
\alpha_{v,w}: f(\underline{\mu},\underline{\lambda}) \otimes M \mapsto
f(\xi^v \underline{\mu}, \xi^{-w} \underline{\lambda}) \otimes E_{v,w} M E_{v,w}^{-1},
\end{equation}
with $\underline{\mu}=(\underline{\mu}_1,\ldots, \underline{\mu}_n)=
(\mu_{11},\ldots, \mu_{1m},\ldots, \mu_{n1},\ldots,\mu_{nm})$ and similarly
for $\underline{\lambda}$, where the notation $\xi^v \underline{\mu}$ means
$\xi^v \underline{\mu}= (\xi^{a_{ij}} \mu_{ij})_{i=1,\ldots,n; j=1,\ldots,m}$, with
$v=(x_1,\ldots,x_n)$ and each $x_i=(a_{i1},\ldots,a_{im})$. The notation 
$\xi^{-w} \underline{\lambda}$ is analogous. 
We realize here the matrix algebra $M_{q^n}(\C)$ as in Lemma \ref{MCpsigma},
as the algebra $C^*(\cE/\cZ)=C^*((\Z/p\Z)^{2mn},\sigma)$ generated by elements
$E_{v,w}$ as in \eqref{Evw}.  

The fixed point algebra of the action \eqref{actalphavw}  defines the 
endomorphism algebra of a vector bundle on the torus
$T^{2r}$ of rank $q^n$. 
The external tensor product $E_1\boxtimes E_2$
of two vector bundles $V_1$ and $V_2$, respectively
over base spaces $X_1$ and $X_2$, is the vector bundle over $X_1\times X_2$ given by
$\pi_1^*(V_1)\otimes \pi_2^*(V_2)$, with $\pi_1$ and $\pi_2$ the projections of $X_1\times X_2$
onto the two factors.  We then see that the vector bundle on $T^{2r}$ described above is, in fact,
the $r$-times external tensor product of the bundle $E_p$ on $T^2$, since the action
\eqref{actalphavw} is the product of an action of the form \eqref{actalpha} on each copy
of $C(T^2,M_p(\C))$. Thus, the fixed point algebra is the algebra of endomorphisms
$\Gamma(T^{2r},E_p^{\boxtimes r})$.

The fixed point algebra of the action \eqref{actalphavw} on $C(T^{2r},M_{q^n}(\C))$
is generated by elements of the form $\underline{\mu}(v)\otimes \underline{\lambda}(w)
\otimes E_{v,w}$, where $\underline{\mu}(v,w)$ is the tensor product of those 
$\underline{\mu}(v)_{ij}$ for which $a_{ij}= 0$, and similarly for $\underline{\lambda}(w)$.
Given the explicit form of the elements $E_{v,w}$ as in \eqref{Evw}, we see that
the fixed point algebra is equivalently generated by elements of the form
$\mu_{ij}\otimes (1\otimes\cdots \otimes T\otimes \cdots 1)$, with $T$ in the $(i,j)$-th
coordinate of the tensor product, and $\lambda_{ij} \otimes (1\otimes\cdots\otimes R \otimes
\cdots \otimes 1)$, with $R$ in the $(i,j)$-th place. Thus, it is the $r$-fold tensor product 
$\cA_{1/p}^{\otimes r}$ of the algebra $\cA_{1/p}$ of the rational noncommutative torus. 

Now suppose one is given a $q$-ary quantum stabilizer code of length $n$ and
size $k$. This means that we have a commutative subgroup $\cS$ of $\cE$ and
a character $\chi:\cS \to U(1)$ that is trivial on $\cZ$ and such that the common
eigenspace $Q_{\cS,\chi}\subset \C^{q^n}$ on which the operators $s\in \cS$
act as $s\psi =\chi(s) \psi$ has complex dimension $k$.

The choice of the commutative subgroup $\cS$ of $\cE$ determines a commutative
subalgebra $\cA_{\cS}$ of the algebra $\cA_{1/p}^{\otimes r}$, which is the subalgebra
generated by elements of the form $\underline{\mu}(v)\otimes \underline{\lambda}(w)
\otimes E_{v,w}$ as above, with $E_{v,w}\in \cS$.  This is the commutative subalgebra
of the endomorphism algebra $\Gamma(T^{2r},E_p^{\boxtimes r})$, generated by the unitaries
$\underline{\mu}(v)\otimes \underline{\lambda}(w)\otimes E_{v,w}$.  

The common eigenspaces of the $E_{v,w}\in \cS$ acting on 
$\C^{q^n}$ correspond to characters $\chi$ of $\cS$. Thus, the eigenspace $Q_{\cS,\chi}$,
for the character $\chi$ of the data of the $q$-ary quantum stabilizer code, determines a
subbundle $\cF_{\cS,\chi}$ of the bundle $E_p^{\boxtimes r}$ over $T^{2r}$ 
with an action of the abelian subalgebra $\cA_{\cS}$ of $\cA_{1/p}^{\otimes r}$ by
endomorphisms. 
\endproof

We can give a more explicit description of the algebra $\cA_\cS$ as follows.

\begin{cor}\label{ASTchi}
The algebra $\cA_{\cS}=C(X_{\cS})$ is the algebra of functions of a
space $X_{\cS}=\cup_{\chi\in \hat\cS} T_\chi$, where $T_\chi$ is a
quotient of the torus $T^{2r}$ over which the bundle $\cF_{\cS,\chi}$
descends to a direct sum $\cL_{\cS,\chi}^{\oplus k}$ of $k$-copies of a line bundle.
\end{cor}

\proof
The abelian subalgebra $\cA_{\cS}$ of
$\cA_{1/p}^{\otimes r}$ can be identified, via the Gelfand--Naimark correspondence,
with the algebra of functions $C(X_{\cS})$ on a compact Hausdorff topological space
$X_{\cS}$.  To give an explicit description of the space $X_{\cS}$ in relation to the
torus $T^{2r}$, it is convenient to also view $\cA_{\cS}$ as the subalgebra of the
abelian algebra $C(T^{2r}, \C[\cS])$ generated by the elements 
$\underline{\mu}(v)\otimes \underline{\lambda}(w)\otimes E_{v,w}$ as
above, with $E_{v,w}\in \cS$.  We write these elements in shorter notation as
$\mu_s \otimes \lambda_s \otimes s$, for $s\in \cS$. For varying $s\in \cS$, 
the corresponding $\mu_s\otimes \lambda_s$ generate a subalgebra 
$C(T^{2r})$, which corresponds to a quotient space of $T^{2r}$. 

By Pontrjagin duality, we can identify $\C[\cS]$, which is the same as $C^*(\cS)$ since
$\cS$ is a finite (abelian) group, with $C(\hat\cS)$, for $\hat\cS$ the character group. 
The isomorphism $C^*(\cS)\simeq C(\hat\cS)$ is by Fourier transform.
Since $\hat\cS$ is also a finite (abelian)
group, $C(\hat\cS)=\oplus_{\chi \in \hat\cS} \C_\chi$, where $\C_\chi$ is the 1-dimensional
algebra of functions on the point $\chi\in \hat\cS$.  Thus, we have 
$C(T^{2r}, \C[\cS])=C(T^{2r}\times \hat\cS)=\oplus_{\chi \in \hat\cS} C(T^{2r})\otimes \C_\chi$. 
The component in $C(T^{2r})\otimes \C_\chi$ of the subalgebra $\cA_\cS$, which we
denote by $\cA_{\cS,\chi}$ is then generated by the elements of the form
$\mu_s \otimes \lambda_s \otimes \hat\delta_s  p_\chi$, where $\hat\delta_s \in C(\hat\cS)$
is the Fourier transform of the generator $\delta_s$ of $\C[\cS]$, and $p_\chi$ is the
projection onto the $\C_\chi$ component of $C(\hat\cS)$, where 
$\hat\delta_s  p_\chi= \chi(s)$.  Upon denoting by $T_\chi$ the 
the quotient space of $T^{2r}$ that corresponds to
the subalgebra of $C(T^{2r})$ generated by the $\mu_s \otimes 
\lambda_s \otimes \hat\delta_s  p_\chi$, we get 
$\cA_\cS=\oplus_{\chi\in \hat\cS} C(T_\chi)\otimes \C_\chi$. 

By construction, the subbundle $\cF_{\cS,\chi}$ then restricts to $T_\chi$ as a direct
sum $\cL_{\cS,\chi}^{\oplus k}$ of $k$-copies of a line bundle $\cL_{\cS,\chi}$, whose
sections transform as $(\mu,\lambda,z)\mapsto
(\mu_s \mu, \lambda_s \lambda, \chi(s) z)$.
\endproof

\subsection{Classical codes and the rational noncommutative torus}

We show next how, in the case of a  quantum stabilizer code
obtained from a self-orthogonal classical linear code via the CSS 
algorithm, one can read some of the properties of the classical
code in the algebra $\cA_{\cS}$.

\smallskip 

Let $C$ be a classical linear code $C\subset \F_q^n$ and let $Q_{\cS_C,\chi}$ be a 
$q$-ary quantum stabilizer code obtained from $C$ via the CSS algorithm recalled above.
Recall that, for a code word $c\in C$ the Hamming weight $\varpi(c)$ is the number of
non-zero coordinates of $c\in\F_q^n$. 

\begin{prop}\label{ASTs}
The algebra $\cA_{\cS}=C(X_{\cS})$ has a natural filtration by the the Hamming
weight of words in the classical code $C$.
\end{prop} 

\proof Seen as a subalgebra of $C(T^{2r})\otimes \C[\cS]$, the commutatie
algebra $\cA_{\cS}$ is generated by elements of the form \
$\mu_s\otimes\lambda_s\otimes \delta_s$, where the $\mu_s$ and $\lambda_s$
are defined as above as the $\mu_{ij}$ and $\lambda_{ij}$,
respectively for the indices $(i,j)$ for which $a_{ij}=0$ and $b_{ij}=0$ in the
coordinates of $(v,w)$, for $s=E_{v,w}\in \cS$. Thus, we can write the algebra
as $\cA_S=  \oplus_{s\in \cS} C(T_s) \otimes \delta_s$, where $C(T_s)$ is the
subalgebra of $C(T^{2r})$ generated by the $\mu_s$ and $\lambda_s$ as above. 
The spaces $T_s$ are quotients of $T^{2r}$ of dimension equal to $2r- \varpi(v,w)$,
where $\varpi(v,w)$ is the Hamming weight of the word $(v,w)$. 
Under multiplication in the algebra, the products of a generator of the form
$\mu_s\otimes \lambda_s\otimes \delta_s$ and a generator of the form $\mu_{s'}\otimes
\lambda_{s'}\otimes \delta_{s'}$ are (strictly) contained among the set of generators of the
form $\mu_{s+s'} \otimes \lambda_{s+s'} \otimes \delta_{s+s'}$, hence $C(T_s)\otimes\delta_s
\cdot C(T_{s'})\otimes \delta_{s'} \subset C(T_{s+s'}) \otimes \delta_{s+s'}$, so that the
filtration by the Hamming weight is compatible with the algebra structure on $\cA_{\cS}$.
\endproof

\section{Algebras and spaces of classical and quantum codes}\label{AlgCQcodeSec}

In this section we modify the previous setting to describe a noncommutative
space where the pairs of a classical linear code and the corresponding 
quantum stabilizer code can be embedded as subspaces in a uniform
way. This is based on a modification of the previous construction, where
the rational noncommutative tori, obtained from endomorphism
algebras of vector bundles over tori, are replaced by spaces
obtained as bundles over tori with fiber a Cantor set.
These are obtained by considering the fractals and the operator algebras 
associated to classical codes as in \cite{ManMar}.

\subsection{Classical codes and fractals}\label{FractalSec}

As shown in \cite{ManMar}, to a classical (not necessarily linear) code $C\subset \fA^n$,
one can associate a fractal $\Lambda_C$ by identifying the alphabet $\fA$ with $\# \fA=q$ with
the digits of the $q$-ary expansion of numbers in the interval $[0,1]$, so that infinite
sequence of code words $x_0 x_1 x_2 \ldots$ determine a subset $\Lambda_C$ of point in 
the cube $[0,1]^n$. This subset is typically a Sierpinski fractal. The parameters of
the code are related to the Hausdorff dimension of $\Lambda_C$ and to the Hausdorff
dimension of its intersections with translates of coordinate hyperplanes (see \cite{ManMar}).

\begin{center}
\begin{figure}
\includegraphics[scale=0.95]{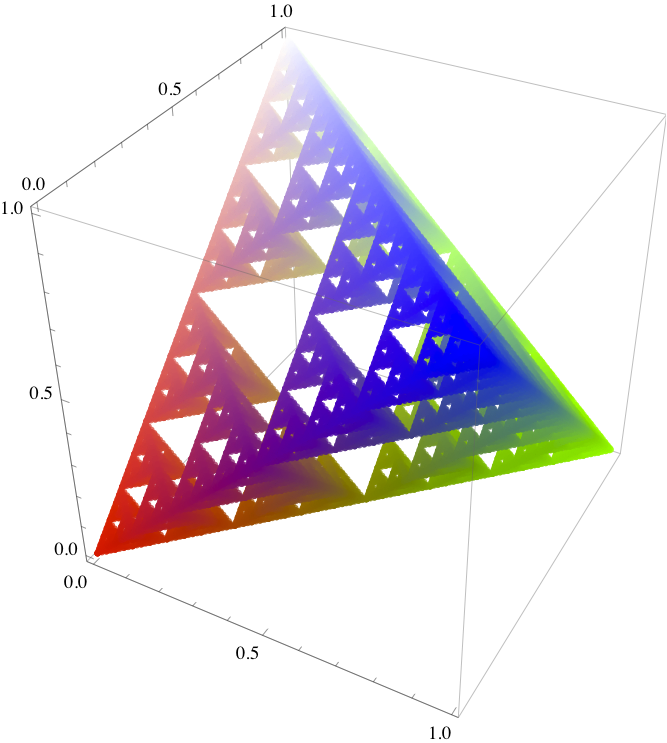}
\caption{The fractal associated to the code $C$ of \eqref{codeexample}. \label{fractalFig}}
\end{figure}
\end{center}

To see concretely the fractal structure associated to a code, 
consider the simple example of the $[3,2,2]_2$ code $C$ given by
\begin{equation}\label{codeexample}
 C = \left\{ \begin{array}{c} (0,0,0) \\ (0,1,1) \\ (1,0,1) \\ (1,1,0) \end{array} \right. 
\end{equation}
In this case, the corresponding fractal is the Sierpinski gasket illustrated in Figure \ref{fractalFig}.

\subsection{Noncommutative spaces and quantum statistical systems from codes}

In \cite{ManMar} it was also suggested to consider operator algebras associated to
a classical code $C\subset \fA^n$, in the form of a Toeplitz algebra $\cT_C$
generated by isometries $S_a$ for $a\in C$, $S_a^* S_a=1$, with mutually 
orthogonal ranges, and the Cuntz algebra $\cO_C$, which is the quotient 
of $\cT_C$ obtained by imposing the additional relation $\sum_{a\in C} S_a S_a^*=1$.
The Cuntz algebra $\cO_C$ has a natural representation as bounded operators
on the Hilbert space $L^2(\Lambda_C, d\mu)$ with the Hausdorff measure of
dimension $\dim_H(\Lambda_C)$, where the generators $S_a$ act as 
$$ S_a f(x) = \chi_{\sigma_a(\Lambda_C)}(x) \Phi_a(\sigma(x))^{-1/2} f(\sigma(x)). $$
Here $x$ is an infinite sequence of code words, $x=(x_1,\ldots, x_n)$ with each 
$x_i=x_{i0}  \cdots x_{in} \cdots$, $x_{ij}\in \fA$, $(x_{1j}, \ldots, x_{nj})\in C$.
The map $\sigma_a$ on $\Lambda_C$ is given by $\sigma_a(x)=(a_1 x_1,\ldots, a_n x_n)$, 
for $a=(a_1,\ldots,a_n)\in C$ and the map $\sigma$ is
the one-sided shift that removes the $(x_{10},\ldots,x_{n0})$ code word of $x$ and
returns the same infinite sequence of code words shifted on step to the left,
starting with $(x_{11}, \ldots, x_{n1})$. The function $\Phi_a$ is the Radon--Nikodym
derivative of the Hausdorff measure, $\Phi_a(x)=d\mu\circ\sigma_a / d\mu$.

As shown in \cite{ManMar}, the natural time evolution on the Toeplitz algebra $\cT_C$
given by $\sigma_t(S_a) = q^{int} S_a$ defines a quantum statistical mechanical
system that has as partition function $Z_C(\beta) = (1-q^{(R-\beta)n})^{-1}$,
with $R$ the rate of the code $C$. This is the same as the structure function of
the language $\Lambda_C$, so that the entropy of the language (which is the log of
the radius of convergence) agrees with the rate of the code.

\subsection{Linear codes and group actions}

In the case of linear codes, one can enrich the construction above with additional
structure.

\smallskip

Let $C\subset \F_q^n$ be a linear code. Let $G_C$ be the additive group
generated by the basis vectors of $C$. Then $G_C$ acts on the algebras
$\cT_C$ and $\cO_C$ by $\gamma_a: S_b \mapsto S_{b+a}$. This action
shuffles the indices of the generating isometries hence it preserves the
relations. Thus, one can consider the algebras $\cT_C \rtimes G_C$
and $\cO_C \rtimes G_C$. These are generated by elements of the form
$S_a \gamma_b$ with product $S_a \gamma_b S_{a'} \gamma_{b'}
=S_a S_{b+a'} \gamma_{b+b'}$. 

\smallskip

Actions of finite abelian groups on Cuntz algebras were studied
extensively in operator algebras, in relation to the Rokhlin property. 

\smallskip

\subsection{The Rokhlin property}

Finite group actions on $C^*$-algebras that have the Rokhlin property have
been widely studied in the context of classification problems for $C^*$-algebras.
The Rokhlin property for an action $\alpha$ of a finite group $G$ on a $C^*$-algebra $A$
prescribes the existence, for any finite $F\subset A$ and any $\epsilon >0$, of 
mutually orthogonal projections $e_g$ in $A$, for $g\in G$, such that 
$\| \alpha_g(e_h)-e_{gh} \|< \epsilon$ for all $g,h\in G$; $\| e_h a- a e_g \| < \epsilon$,
for all $g\in G$ and $a\in F$, and $\sum_{g\in G} e_g=1$.
The importance of the Rokhlin property lies in the fact that it ensures that the
group actions are classifiable in terms of $K$-theoretic invariants.
The case of quasi-free actions of finite groups on Cuntz algebras was 
considered in \cite{Izumi}.

\begin{lem}\label{GCRok}
The action of $G_C$ on the Cuntz algebra $\cO_C$ has the Rokhlin property.
\end{lem}

\proof
According to \cite{Izumi}, an action $\alpha$ of a topological group $G$
on the Cuntz algebra $\cO_n$ is quasi-free if $\alpha_g$ globally preserves
the linear span $\cH_n$ of the generators $\{ S_i \}_{i=1,\ldots, n}$ of the
Cuntz algebra, for each $g\in G$. 
The action of $G_C$ on $\cO_C$ described above is
quasi-free in this sense, since it has the effect of permuting the
generators $S_a$ of $\cO_C$, so it leaves the corresponding space, which we
denote by $\cH_C$, invariant. 
One then sees directly from Proposition 5.6 and Example 5.7 of \cite{Izumi}, that 
the action of $G_C$ on $\cO_C$ has the Rokhlin property. 
\endproof

We also mention here that, according to Proposition 5.5 of \cite{Phill}, 
an action of a finite group $G$ on a Cantor set has the Rokhlin property 
if and only if the action is free. Later in this section we relate the
action of $G_C$ on $\cO_C$ to an action on the fractal $\Lambda_C$.

\subsection{Twisted crossed products and codes}

One can twist the crossed product algebras $\cT_C \rtimes G_C$ and
$\cO_C \rtimes G_C$ by the cocycle $\sigma$ as in \eqref{Matsigma}.

\begin{lem}\label{TGsigma}
Let $C \subset \F_{p^{2m}}^n$ be a linear code with $\# C=q^k$, with $q=p^{2m}$. 
Then $G_C\subset (\Z/p\Z)^{2mn}$ is $G_C\simeq (\Z/p\Z)^{2mk}$
and the multiplier \eqref{Matsigma} defines twisted crossed product algebras
$\cT_C \rtimes_\sigma G_C$ and $\cO_C \rtimes_\sigma G_C$. 
\end{lem}

\proof The twisted crossed product algebras are generated by elements 
$S_{(a,b)} \gamma^\sigma_{(v,w)}$ with $(a,b)\in C$ and $(v,w)\in (\Z/p\Z)^{2mk}$,
with the product given by
$$ S_{(a,b)} \gamma^\sigma_{(v,w)} S_{(a',b')} \gamma^\sigma_{(v',w')} =
\sigma(v,v')  S_{(a,b)} S_{(v+a',w+b')} \gamma^\sigma_{(v+v',w+w')} . $$
The associativity, as above, is ensured by the multiplier properties of $\sigma$.
\endproof

\begin{lem}\label{Lambdacrossprod}
The (twisted) action of $G_C$ on $\cO_C$ preserves the maximal abelian subalgebra
of $\cO_C$ isomorphic to $C(\Lambda_C)$. 
\end{lem}

\proof The action of $G_C$ on the generators $S_a$ of $\cO_C$ is given by
$\gamma_b S_a = S_{a+b}$. The subalgebra of $\cO_C$ isomorphic to $C(\Lambda_C)$
is generated by the range projections $S_\alpha S_\alpha^*$, where $S_\alpha$, for some
multi-index  $\alpha=(a_1,\ldots, a_m)$, $a_i\in C$, 
is a finite product $S_\alpha =S_{a_1}\cdots S_{a_m}$
of generators.  The range projection $S_\alpha S_\alpha^*$ corresponds to the projection
in $C(\Lambda_C)$ given by the characteristic function of the subset $\Lambda_C(\alpha)$
of infinite sequences of code words in $\Lambda_C$ that start with the word $\alpha$.

The induced action $\gamma$ of the group $G_C$ on the 
fractal $\Lambda_C$ is then determined
by the action on $C(\Lambda_C)$ that maps the characteristic function 
$\chi_{\Lambda_C(\alpha)}= S_\alpha S_\alpha^*$ to the characteristic function
$\chi_{\Lambda_C(\gamma_b(\alpha))}=\gamma_b(S_{\alpha})\gamma_b(S_{\alpha}^*)$,
where $\gamma_b(S_\alpha) =\gamma_b(S_{a_1})\cdots \gamma_b(S_{a_m})=
S_{a_1+b}\cdots S_{a_m+b}$. 

This implies that the induced action on the Cantor set is given by addition in each
digit of the expansion: for $(x,y)\in \Lambda_C$ given by
$(x,y)=(x_0 x_1\ldots x_N \ldots, y_0 y_1\ldots y_N \ldots)$ with $(x_i,y_i)\in C$, one gets 
$\gamma_{v,w}(x,y) =((x_0+v)(x_1+v)\ldots (x_N+v)\ldots, (y_0+w)(y_1+w)\ldots
(y_N+w)\ldots)$, with $(x_i+v,y_i+w)\in C$. 

Thus, one obtains a subalgebra
$C(\Lambda_C) \rtimes_\sigma G_C$ of $\cO_C\rtimes_\sigma G_C$ of
the twisted crossed product. Elements of this subalgebra can be written as
\begin{equation}\label{aCLambdaG}
a = \sum_{(v,w)\in C} f_{(v,w)}(x,y) \, \gamma_{(v,w)}^\sigma,
\end{equation}
for $f_{(v,w)}\in C(\Lambda_C)$ and $\gamma_{(v,w)}^\sigma$ as above, with
$$ f_{(v,w)}(x,y) \gamma_{(v,w)}^\sigma f_{(v',w')}(x,y) \gamma_{(v',w')}^\sigma = $$ $$
\sigma((v,w),(v',w')) \,  f_{(v,w)}(x,y)  f_{(v',w')}(\alpha_{v,w}(x,y))\, 
\gamma_{(v+v',w+w')}^\sigma . $$
\endproof

Consider then a quantum stabilizer code $Q=Q_{\cS,\chi}$, associated
to a classical self-orthogonal linear code $C$ in $\F_p^{2nm}$, with
an $\F_p$-automorphism $\varphi\in \Aut(\F^m_p)$,
so that $\cS=\{ \xi^k E_{v,\varphi(w)} \,|\, (v,w) \in C \}$ is an abelian
subgroup of $\cE$. Thus, $Q=Q_{C,\varphi}$. 
Because of the self-orthogonal condition, the cocycle $\sigma((v,w),(v',w'))
=\xi^{ -\langle w,v' \rangle}$ is trivial, so the crossed product algebras 
$\cO_C \rtimes_\sigma G_C$ and $C(\Lambda_C)\rtimes_\sigma G_C$ are
just the untwisted $\cO_C \rtimes G_C$ and $C(\Lambda_C)\rtimes G_C$
with $G_C$ the abelian group identified with the subgroup of $\cS\subset \cE$
with elements the $E_{v,\varphi(w)}$. The same holds for the related algebras
$\cT_C \rtimes_\sigma G_C$ which is $\cT_C \rtimes G_C$.

\subsection{Adding machines}

The action of $G_C$ on $\Lambda_C$ induced by the action on $\cO_C$
is simply the coordinate-wise translation on the linear code $C$, when we 
identify the fractal $\Lambda_C$ with an infinite product of copies of $C$.
There is another, more interesting way in which one can use the linear
structure of the code $C$ to directly construct an action of $G_C$ on $\Lambda_C$,
which is better behaved from the dynamical systems point of view, namely
an odometer action or adding machine.

This is modeled on the action of $\Z$ on a Cantor set $X$ which is an infinite 
product of cyclic groups, $X=\prod_k \Z/n_k \Z$, given by addition by one
with carry to the right, namely
$$ T (x_0,x_1,x_2,\cdots, x_k, \cdots) =\left\{ \begin{array}{ll} 
(x_0+1, x_1, x_2, \cdots, x_k, \cdots) & 
x_0\neq n_0-1 \\[2mm] (0,x_1+1,x_2,\cdots, x_k,\cdots) & x_0=n_0-1, x_1\neq n_1-1 \\[2mm]
\vdots & \vdots \\[2mm]
(0,0,0,0,\ldots, 0,\ldots) & x_k=n_k-1, \forall k. \end{array}\right. $$

Let us first see that on an explicit example. 
We consider again the fractal $\Lambda_C$ of Figure \ref{fractalFig}, 
for the code \eqref{codeexample}. In that case we can identify the code $C$
with the 2-dimensional $\F_2$-vector space generated by the vectors
$e_1=(0,1,1)$ and $e_2=(1,0,1)$, and consisting of the vectors
$\{ 0, e_1,e_2,e_3=e_1+e_2 \}$ with $e_3=(1,1,0)$. Then, we can 
write code words $c\in C$ as $c=x e_1 + y e_2$ with $x$ and $y$ in $\Z/2\Z$.
Thus, we write elements in $\Lambda_C$ as infinite sequences
$(x_0,x_1,x_2,\ldots,x_k,\ldots,y_0,y_1,\ldots,y_k,\ldots)$ with the
$x_i$ and $y_i$ in $\Z/2\Z$ and $c_i=x_i e_1 + y_i e_2$ in $C$.
The group $G_C=(\Z/2\Z)^2$ then defines an odometer map by setting
$T_{(1,0)} (x_0,x_1,x_2,\ldots,x_k,\ldots,y_0,y_1,\ldots,y_k,\ldots)$ to be equal to
$(x_0+1,x_1,\ldots,x_k,\ldots,y_0,y_1,\ldots,y_k,\ldots)$ if $x_0\neq 1$, to
$(0,x_1+1,\ldots,x_k,\ldots,y_0,y_1,\ldots,y_k,\ldots)$ if $x_0=1$ and $x_1\neq 1$,
and so on, and similarly for $T_{(0,1)}$ and the $y_i$ coordinates.

More precisely, in the general case of a linear code $C\subset \F_p^{2mn}$, with $\# C=q^k$
and $q=p^{2m}$, we identify the fractal $\Lambda_C$
with an infinite product of copies of the code $C$, with the topology generated by the
cylinder sets $\Lambda_C(\alpha)$ as above. 
Upon choosing a basis, we can identify the linear code with $C\simeq (\Z/p\Z)^{2mk}$, 
so that we can write code words $(v,w)\in C$ as vectors $(v,w)=(v_{ij},w_{ij})$ with
$i=1,\ldots,m$ and $j=1,\ldots,k$ and $v_{ij}$ and $w_{ij}$ in $\Z/p\Z$.  
The group $G_C$ is then itself identified with $G_C\simeq (\Z/p\Z)^{2mk}$ and it
defines an adding machine on $\Lambda_C$ by setting 
$T_{(1_{ij},0)} (x_0, x_1,\cdots, x_k,\cdots, 
y_0,y_1,\cdots, y_k,\cdots)$ to be 
$$ \left\{ \begin{array}{ll} (x_0+1_{ij},x_1,\cdots,x_k,\cdots,y_0,y_1,\cdots,y_k,\cdots) &
(x_0)_{ij} \neq p-1  \\[2mm]
((\hat x_0)_{ij} ,x_1+1_{ij},\cdots,x_k,\cdots,y_0,y_1,\cdots,y_k,\cdots) &
(x_0)_{ij}=p-1, (x_1)_{ij}\neq p-1 \\[2mm]
\vdots & \vdots
\end{array}\right. $$ 
where $(\hat x_0)_{ij}$ means the vector that has a zero at the $(i,j)$-th
coordinates and all the other coordinates equal to those of $x_0$, 
and similarly for $T_{(0,1_{ij})}$ on the $(y_k)_{ij}$ coordinates. 

One can then consider the crossed product of the algebra of functions
$C(\Lambda_C)$ by this odometer action of the group $G_C$. This
gives a different kind of operator algebra, more in the style of a
Bunce--Deddens algebra, that can be used to study properties of the
code $C$. Although we do not pursue this line extensively in this paper,
we make some further comments about it in the final section.

\subsection{Disconnection and group actions}

Consider points of $T^2=S^1 \times S^1$ as points in the square
$Q^2=[0,1]\times [0,1]$ with the boundary identifications that give $T^2$,
where we write the points of $[0,1]$ in terms of their $p$-ary digital
expansion: $x=0.x_1 x_2 x_3 \ldots x_N \ldots$, with $x_i \in \{ 0, \ldots, p-1\}$.
As in the decimal case, the expansion is a $1:1$ representation on the 
irrational points and $2:1$ on the rational points.  Fixing the first $N$
digits of the expansion determines a subinterval of $[0,1]$ of length
$p^{-N}$. 

\smallskip

There is a totally disconnected compact topological space $T^2_\Q$, called
the disconnection of $T^2$ at the rational points, which maps surjectively 
to $T^2$ with a map that is $1:1$ over the irrational points and $2:1$ over the
rational points. As a topological space, it is the spectrum of a commutative
$C^*$-algebra $C(T^2_\Q)$, which is the smallest $C^*$-algebra containing
$C(T^2)$ in which all the characteristic functions of intervals $[k p^{-N},(k+1)p^{-N})$ 
with $k\in \{ 0, \ldots, p-1 \}$ and $N\geq 1$ are continuous functions. 

\smallskip

\begin{lem}\label{actT2Q}
The group $(\Z/p\Z)^2$ acts on the disconnection $T^2_\Q$ by 
\begin{equation}\label{actklT2Q}
\gamma_{(k,\ell)}(x,y)=(\gamma_k(x_0) \gamma_k(x_1)\ldots \gamma_k(x_N)\ldots,
\gamma_\ell(y_0) \gamma_\ell(y_1) \ldots \gamma_\ell(y_N) \ldots), 
\end{equation}
where, for $a \in \Z/p\Z$, $\gamma_b(a)=a+b$ in $\Z/p\Z$. One can then form a
crossed product algebra $C(T^2_\Q)\rtimes_\sigma (\Z/p\Z)^2$, with the action
\eqref{actklT2Q}, and with the twisting given by 
the cocycle $\sigma((v,w),(v',w'))=\xi^{ -\langle w,v' \rangle}$.
\end{lem}

\proof The action $(x_i,y_i)\mapsto (\gamma_k(x_i),\gamma_\ell(y_i))$
on the $i$-th digit of the p-ary expansion of $(x,y)\in T^2_\Q$ has the
effect of moving the product of intervals $[x_i p^{-i}, (x_i+1) p^{-i})\times [y_i p^{-i}, (y_i +1) p^{-i})$
inside $T^2$ to $[(x_i +k \mod p) p^{-i}, (x_i +1 +k \mod p) p^{-i})\times 
[(y_i +k \mod p) p^{-i}, (y_i +1 +k \mod p) p^{-i})$. While this is not a
continuous function on $T^2$ it becomes continuous on the totally disconnected $T^2_\Q$.
Thus, one can form the crossed product $C^*$-algebra with respect to this action.
It is generated by elements of the form $\sum_{g \in (\Z/p\Z)^2} h_g(\lambda,\mu)
r^\sigma_g$, with $(\lambda,\mu)\in T^2_\Q$ and where
$r^\sigma_{g_1} r^\sigma_{g_2} = \sigma(g_1,g_2) r^\sigma_{g_1 g_2}$ 
and $r^\sigma_g h(\lambda,\mu) = h(\gamma_g(\lambda,\mu)) r^{\sigma}_g$.
\endproof

Notice that we can also, as above, consider the adding machine action
on $T^2_\Q$ and proceed in a similar way.

\medskip

\subsection{Cantor set bundles}

We start with the geometric setting we have discussed above in \S \ref{QtoriSec}
and we see how that gets modified when we also take into account the 
fractal geometry $\Lambda_C$ associated to the classical code $C$.

\smallskip

We have seen that a q-ary quantum stabilizer code $Q_{\cS,\chi}$ of length $n$ and size $k$ 
identifies a commutative subalgebra $\cA_\cS$ of the endomorphism algebra 
$\Gamma(T^{2r}, {\rm End}(E_p^{\boxtimes r}))$ of a vector bundle $E_p^{\boxtimes r}$
over the torus $T^{2r}$, where $q=p^m$ and $r=nm$. 

\smallskip

\begin{prop}\label{fibrLambda}
If $C\subset F_q^{2n}$ is a self-orthogonal linear code and $Q_{\cS_C,\chi}$
the associated q-ary quantum code, the fractal $\Lambda_C$ can
be embedded in the disconnection $T^{2r}_\Q$. The pullback of the
subbundle $\cF_{\cS,\chi}\subset E_p^{\boxtimes r}$ to $\Lambda_C$ 
via the projection $T^{2r}_\Q\to T^{2r}$ and its quotient by the action of $\cS_C$
determine a fibration over a torus with fiber $\Lambda_C$.
\end{prop}

\proof
We can pull back the bundle $E_p^{\boxtimes r}$ along the projection
map $\pi: T^{2r}_\Q \to T^{2r}$ and further restrict it to $\Lambda_C$
by pulling it back along the embedding $\iota: \Lambda_C \hookrightarrow T^{2r}_\Q$. 

In fact, the fractal $\Lambda_C$ can be realized 
as a subspace of the product $(T^2_\Q)^n$, by identifying points of $\Lambda_C$, which
are infinite sequences of code words $c=c_1 c_2 \ldots c_N \ldots$, with $c_i \in C 
\subset \F_q^{2n}\simeq \F_p^{2r}$, with points of $(T^2_\Q)^r$, by writing each $c_i$ as a pair
of $r$-tuples of elements in $\Z/p\Z$, $c_i=(x_{i,1},\ldots , x_{i,r}, y_{i,1},\ldots, y_{i,r})$,
hence identifying the pair $(x_j,y_j)$ of sequences $x_j=x_{1,j} x_{2,j} \ldots x_{N,j}\ldots$ 
and $y_j=y_{1,j} y_{2,j}\ldots y_{N,j} \ldots$, $j=1,\ldots,n$ with the p-ary expansion 
of a point in $T^2_\Q$, hence $(x,y)\in (T^2_\Q)^n$, with $(x,y)=(x_1,\ldots,x_n,y_1,\ldots,y_n)$.

Over $\Lambda_C$ the induced vector bundle can be trivialized, so that
$\iota^* \pi^* E_p^{\boxtimes r} \simeq \Lambda_C \times \C^{q^n}$. The subbundle
$\cF_{\cS,\chi}$ of  $E_p^{\boxtimes r}$ identified by the q-ary quantum stabilizer code 
$Q_{\cS,\chi}$ in turn pulls back to a subbundle $\iota^* \pi^* \cF_{\cS,\chi} \simeq 
\Lambda_C \times Q_{\cS,\chi}$. 

We now assume that $C$ is a self-orthogonal linear code and that $Q_{\cS,\chi}$
is the associated q-ary quantum code, under the CSS algorithm. When we take
into account the action of $G_C$ on the linear code $C$, we then have compatible
actions
$$\xymatrix{ \iota^* \pi^* E_p^{\boxtimes r}   \ar@{->}[r]^{\Phi_{(v,w)}} \ar@{->}[d]
& \iota^* \pi^* E_p^{\boxtimes r}  \ar@{->}[d] \\
\Lambda_C  \ar@{->}[r]^{\gamma_{(v,w)}} & \Lambda_C } $$
where in the trivialization $\iota^* \pi^* E_p^{\boxtimes r} \simeq \Lambda_C \times \C^{q^n}$,
the action on $\iota^* \pi^* E_p^{\boxtimes r}$ is given by 
$\Phi_{(v,w)}=(\gamma_{(v,w)},E_{v,w})$. The action preserves the subbundle $\cF_{\cS,\chi}$,
where the induced action is through the character $\chi$,
$$ \Phi_{(v,w)}=(\gamma_{(v,w)},\chi(E_{v,w})). $$

When taking the quotient with respect to this action, using the trivializations of the
bundles, one obtains quotient spaces, respectively of the form 
$(\Lambda_C \times \C^{q^n})/\cS_C$ and $(\Lambda_C \times Q_{\cS_C,\chi})/\cS_C$. 
These are, respectively, locally trivial fibrations over the quotients $\C^{q^n}/\cS_C$
and $Q_{\cS_C,\chi}/\cS_C$. We focus in particular on the case of the
subspace $Q_{\cS_C,\chi}$. Because the quotient $Q_{\cS_C,\chi}/\cS_C$ is
singular at the origin, it is preferable to remove this singular point and consider
instead the quotient of $Q_{\cS_C,\chi}^*:=Q_{\cS_C,\chi}\smallsetminus\{ 0\}$.
The action of $\cS_C$ is through the character $\chi$, that is, as multiplication by 
$\chi(E_{v,w}) \in U(1)\subset \C^*$.  Thus, one can further restrict to the unit
vectors and obtain an action on a torus $T^{p^{nm-r}}$, with quotient still
topologically a torus.  The fibration then induced a fibration over this torus 
with fiber a fractal $\Lambda_C$. 
\endproof

Variants of this construction may be useful to better take into account the
dynamical properties of the action of $G_C$ on the fractal $\Lambda_C$.
We give another example below.

\medskip
\subsection{Crossed product algebras and embeddings}

One can also use the fact that the fractal $\Lambda_C$ embeds inside
the disconnection $T^{2r}_\Q$, in a way that is compatible with the
action of $G_C$, to compare different crossed product algebras
$C(\Lambda_C)\rtimes_\sigma G_C$ for different codes inside a
common noncommutative space.

\smallskip

\begin{lem}\label{LambdaT2mor}
Let $\cA=C(T^2_\Q)\rtimes_\sigma (\Z/p\Z)^2$ be the twisted
crossed product algebra of the action of $(\Z/p\Z)^2$ on the 
disconnection $T^2_\Q$. For any classical linear code 
$C\subset \F_p^{2n}$, there is an algebra homomorphism 
$\cA^{\otimes n} \to C(\Lambda_C)\rtimes_\sigma G_C$.
\end{lem}

\proof For $\# C =p^{2k}$, we have $G_C \simeq (\Z/p\Z)^{2k}$.
We regard this as a subgroup $G_C \subset (\Z/p\Z)^{2n}$ of the group of translations
of the whole space $\F_p^{2n}$, as the subgroup of translations that preserve the linear 
subspace $C$. The embedding $\Lambda_C \hookrightarrow T^{2n}_\Q$ described
in Proposition \ref{fibrLambda}
determines an algebra homomorphism $C(T^2_\Q)^{\otimes n} \to C(\Lambda_C)$
given by restriction of functions to $\Lambda_C$.

We write $\alpha: G_C \hookrightarrow
(\Z/p\Z)^{2n}$ for the embedding as a subgroup and $\rho: C((T^2_\Q)^n)\to C(\Lambda_C)$
for the restriction of functions $\rho(f)(x)=f(\iota(x))$, with 
$\iota: \Lambda_C \hookrightarrow (T^2_\Q)^n$ the embedding of the fractal $\Lambda_C$
in the disconnection $T^{2n}_\Q$.
The algebra homomorphism $\rho: C(T^2_\Q)^{\otimes n} \to C(\Lambda_C)$ is compatible
with the action of translations, since we have 
$\gamma_{\alpha(a)} (\iota(x)) = \iota(\gamma_a(x))$, 
for all $x\in \Lambda_C$ and all $a\in G_C$. Thus, we have a morphism of the
crossed product algebras $C(T^{2n}_\Q)\rtimes_\sigma (\Z/p\Z)^{2n} \to
C(\Lambda_C)\rtimes_\sigma G_C$. Finally, we identify 
$C(T^{2n}_\Q)\rtimes_\sigma (\Z/p\Z)^{2n}$
with the tensor product $(C(T^2_\Q)\rtimes_\sigma (\Z/p\Z)^2)^{\otimes n}$.
\endproof

The algebra homomorphisms $\cA^{\otimes n} \to C(\Lambda_C)\rtimes_\sigma G_C$
are constructed as restriction maps, hence in terms of noncommutative spaces
these correspond to embedding the noncommutative spaces associated to
linear codes, whose algebras of coordinates are the $C(\Lambda_C)\rtimes_\sigma G_C$,
into a common noncommutative space, whose algebra of coordinates is $\cA^{\otimes n}$.
The latter therefore can be thought of as a ``universal family" for all the noncommutative
spaces of linear codes $C \subset \F_p^{2n}$, where the total space corresponds to the
``largest" code, namely $\F_p^{2n}$ itself, acted upon by all the translations $(\Z/p\Z)^{2n}$.
Moreover, the subfractals $\Lambda_{C,\ell,\pi}$ associated to linear subcodes $C_\pi$,
which we discuss in the next subsection, determine further compatible specialization maps
$C(\Lambda_C)\rtimes_\sigma G_C \to C(\Lambda_{C_\pi})\rtimes_\sigma G_{C_\pi}$.

\subsection{Minimum distance, subfractals and the weight polynomial}\label{weightpolySec}

We conclude this section with an observation on how one can reinterpret
the weight polynomial of a linear code in terms of subfractals of the 
code fractal, satisfying certain scaling (self-similarity) properties, or
equivalently in terms of counting embeddings to associated Toeplitz
algebras.

\smallskip

We first recall briefly the interpretation of the minimum distance $d$ of a
code $C$ in terms of the fractal geometry of $\Lambda_C$, as given in \cite{ManMar}.
Notice that here we use a slightly different notation from \cite{ManMar} and
our $\Lambda_C$ is the $\bar S_C$ of \cite{ManMar}, so the statement is
slightly different from the one formulated for $S_C$ in that paper, and we 
write it out here explicitly for convenience.

\smallskip

For $\ell=1,\ldots,d$, let $\Pi_\ell$ be the set of $\ell$-dimensional subspaces in $\R^n$
defined by intersections of $n-\ell$ hyperplanes, each of which is a translate of a
coordinate hyperplane. For any given such linear space $\pi\in \Pi_\ell$, we denote by
$\Lambda_{C,\ell,\pi}=\Lambda_C \cap \pi$. The geometry of this intersection 
varies with the choice of the linear space. When non-empty, its form changes 
drastically when $\ell$ increases. More precisely, one has the following (\cite{ManMar}).

\begin{lem}\label{mindfrac}
Let $C\subset \fA^n$ be a code with minimum distance $d=\min\{ d(x,y)\,|\, x\neq y \in C \}$,
in the Hamming metric. For all $\ell< d$, the set $\Lambda_{C,\ell,\pi}$ has 
$\dim_H(\Lambda_{C,\ell,\pi})=0$ and is either empty or it consists of a single point, 
while for $\ell \geq d$ the set $\Lambda_{C,\ell,\pi}$, when non-empty, has an actual fractal structure of positive Hausdorff dimension.
\end{lem}

\proof
The property that $C$ has minimum distance $d$ means that any pair of distinct
points $x\neq y$ in $C$ must have at least $d$ coordinates that do not coincide, since
$d(x,y)=\#\{ i\,|\, x_i\neq y_i \}$. Thus, in particular, this means that no two points of the
code lie on the same $\pi$, for any $\pi$ as above of dimension $\ell \leq d-1$, while
there exist at least one $\pi$ in $\Pi_d$ which contains at least two points of $C$. In terms 
of the iterative construction of the fractal $S_C$, this means the following. For a given 
$\pi\in \Pi_\ell$ with $\ell\leq d-1$, if the intersection $C\cap \pi$ is non-empty it must
consist of a single point. Thus, when restricted to a
linear space $\pi\in \Pi_\ell$ with $\ell\leq d-1$, at the first step the induced
construction of $\Lambda_{C,\ell,\pi}$ consists of replacing the single unit cube of
dimension $\ell$, $Q^\ell =Q^n\cap \pi$, 
with a single copy of a scaled cube of volume $q^{-\ell}$, successively iterating the same
procedure. This produces a single family of nested cubes of volumes $q^{-\ell N}$ with
intersection a single vertex point. The Hausdorff dimension is clearly zero.
When $\ell =d$ one knows there exists a choice of $\pi\in \Pi_d$ for which 
$C\cap \pi$ contains at least two points. Then the induced iterative construction of the
set $\Lambda_{C,\ell,\pi}$ starts by replacing the cube $Q^d =Q^n\cap \pi$ with $\# (C\cap \pi)$
copies of the same cube scaled down to have volume $q^{-d}$. The construction is then
iterated inside all the resulting $\#(C\cap \pi)$ cubes. Thus,  one obtains a set 
of positive Hausdorff dimension $\dim_H(\Lambda_{C,\ell,\pi})$, since we
have a positive solution $s>0$ to the scaling equation $\#(C\cap \pi)\cdot q^{-\ell s}=1$.
\endproof

Thus, as observed in \cite{ManMar},
the parameter $d$ of the code $C$ can be regarded as the threshold value of $\ell$
where the sets $\Lambda_{C,\ell,\pi}$ jump from being trivial to being genuinely fractal objects.

\smallskip

For example, consider the code $C$ of Figure \ref{fractalFig} and \eqref{codeexample}.
The translates of coordinate hyperplanes intersect $C$ in the following way: $C \cap \{x_1=0\}=\{ (0,0,0), (0,1,1) \}$, $C\cap \{ x_1=1 \}=\{ (1,0,1), (1,1,0) \}$, $C\cap \{ x_2=0 \} =
\{ (0,0,0), (1,0,1) \}$, $C\cap \{ x_2=1\} =\{ (0,1,1), (1,1,0) \}$, $C\cap \{ x_3=0 \}=
\{ (0,0,0), (1,1,0) \}$ and $C\cap \{ x_3=1 \} =\{ (0,1,1), (1,0,1) \}$, so that all the
corresponding $\Lambda_{C,2,\pi}$ have positive Hausdorff dimension. On the
other hand, for $\ell=1$, all the intersections of $C$ with an intersection of two
of the above hyperplanes consist of at most one point. 

\smallskip

In the case of linear codes, the Hamming distance
$d(x,y)=\# \{ i \,|\, a_i\neq b_i \}=\#\{ i \,|\, a_i-b_i \neq 0\} =d(x-y,0)$, so that the
minimum distance is measured by $d(C)=\min\{ d(x,0) \,|\, x\in C, \, x\neq 0 \}$. The 
Hamming weight of $x\in C$ is the number of non-zero components of $x$.
Thus, the minimum distance is also the minimum Hamming weight,
$d(C) = \min \{ w(x)\,|\, x\in C, \, x\neq 0 \}$.

\smallskip

Thus, to describe the minimum distance as in Lemma \ref{mindfrac}, 
it suffices to consider those $\pi \in \Pi_\ell$ that are intersections of
coordinate hyperplanes, hence $\F_q$-linear subspaces in $\F_q^n$,
instead of considering also their translates. This identifies subfractals
$\Lambda_{C,\ell,\pi}$ associated to $C_\pi=C\cap \pi$, where the
$C_\pi$ are also linear codes. We write $\Pi_\ell^0\subset \Pi_\ell$
for the set of linear subspaces $\pi$ given by intersections of $\ell$ 
coordinate hyperplanes.

\smallskip

In the example of \eqref{codeexample}, there are three such subfractals
for $\ell=d=2$, which correspond to the intersections with the three
coordinate hyperplanes, $C_1=\{ (0,0,0), (0,1,1) \}$, 
$C_2 =\{ (0,0,0), (1,0,1) \}$, and $C_3 =
\{ (0,0,0), (1,1,0) \}$.

\smallskip

The Toeplitz algebras $\cT_C$ are functorial with respect to
injective maps of sets $f: C \to C'$, with the corresponding morphism of 
algebras mapping $S_a \mapsto S_{f(a)}$. The Cuntz algebras  
are only functorial with respect to bijections. 

\smallskip

Thus, for each set $\Lambda_{C,\ell,\pi}$ of positive Hausdorff dimension, corresponding
to an intersection $C_\pi = C\cap \pi$ with $\# (C\cap \pi) >1$, we have an injective morphism of the
corresponding Toeplitz algebras $T_\pi: \cT_{C_\pi} \to \cT_C$ associated to the inclusion
$C_\pi \subset C$. Moreover, if $\pi$ and $\pi'$ are two elements in $\Pi_\ell$, 
with $\ell\geq d$, such that $\# C_\pi = \# C_{\pi'} >1$, we have an isomorphism of the
corresponding algebras $\cT_{C_\pi} \simeq \cT_{C_{\pi'}}$. 

\smallskip

In the example of $[3,2,2]_2$ code of \eqref{codeexample}, the algebras $\cT_{C_\pi}$ 
for all the translates of the coordinate hyperplanes $\pi  \in \Pi_2$ are isomorphic, and
one correspondingly has six different embeddings of this as a subalgebra of $\cT_C$.
While, if one counts only those that also correspond to linear codes, one has only
three, coming from the intersections of $C$ with the three coordinate hyperplanes,
as above.

\smallskip

For a linear code $C$, one can consider the associated {\em
weight polynomial} of the code $C$. We recall here briefly the 
definition and properties, see \cite{Barg}.
The basic observation is that, for a linear code,  The weight polynomial is given by
\begin{equation}\label{wpoly}
\cA(x,y)= \sum_{i=1}^n \cA_i x^{n-i} y^i , \ \ \ \text{ with } \ \ \ 
\cA_i = \# \{ x\in C\,|\, w(x) =i \}.
\end{equation}
In the example of the code $C$ of \eqref{codeexample}, the weight polynomial
is $\cA(x,y)=x^3 + 3 x y^2$. 

\smallskip

One can then easily see the following interpretation of the coefficients
of the weight polynomial.

\begin{lem}\label{weightembed}
For a linear code $C$, the coefficient $\cA_i$ of the weight polynomial $\cA(x,y)$
is given by 
$$ \cA_i =\# \cup_{ \pi \in \Pi^0_{n-i} } (C_\pi \smallsetminus \{ 0 \}). $$
These linear subcodes $C_\pi$ correspond to subfractals $\Lambda_{C,n-i,\pi}$ 
of $\Lambda_C$ with scaling equation $\# (C\cap \pi) q^{-(n-i)s}=1$.
\end{lem}

\proof Any point $x\in C$ with $w(x)=i$ lies on an intersection of coordinate 
hyperplanes $\pi \in \Pi_{n-i}^0$. Thus, $\cA_i$ counts the number of nonzero
$x\in C$ that lie in some $\pi \in \Pi_{n-i}^0$, that is,
$\cA_i = \# \{ x \neq 0 \in C \,|\, \exists \pi \in \Pi_{n-i}^0 : x\in \pi \}
=\# \{ x \neq 0 \in C \,|\, x \in cup_{\Pi_{n-i}^0} \pi \}$. 
Moreover, if $w(x)=i$ so that $x\in \pi$, for some $\pi \in  \Pi_{n-i}^0$, the
intersection $C_\pi$ is not contained in any $\pi' \in \Pi_{n-i-1}^0$, since $x\notin \pi'$,
so that $\Lambda_{C,n-i,\pi}$ is obtained by scaling $\# C_\pi$ copies of the
cube $Q^{n-i}$ of volume $q^{-(n-i)}$, so that the scaling equation is as stated.
\endproof

Thus, one can view the weight polynomial of the code as a generating function
for the multiplicities of the embeddings $\cT_{C_\pi} \to \cT_C$ for linear subcodes
with $\pi\in \Pi^0_\ell$ giving rise to nontrivial subfractals.

\smallskip

As seen in \cite{ManMar} the Toeplitz algebra $\cT_C$ and the Cuntz algebra $\cO_C$
associated to a classical code $C$ have representations on the Hilbert space 
$L^2(\Lambda_C,d\mu_H)$ and a time evolution $\sigma_t(S_a)=q^{itn} S_a$, 
whose critical temperature KMS state recovers intergration in the Hausdorff 
measure of dimension $\dim_H(\Lambda_C)$ on the fractal $\Lambda_C$.
The embeddings $\cT_{C_\pi} \to \cT_C$ therefore inherit an action  on the
same Hilbert space and the induced time evolution. The critical temperature
KMS state for the time evolution on the subalgebra then recovers the integration
in the Hausdorff measure of dimension $\dim_H(\Lambda_{C,\ell,\pi})$ 
on the subfractal $\Lambda_{C,\ell,\pi}$.

\section{Further remarks and directions}\label{RemkSec}

We have discussed in the previous sections different geometric constructions
that associate to a pair of a classical self-orthogonal linear code $C\subset \F_q^n$ 
and a $q$-ary quantum stabilizer code $Q_{\cS,\chi}$ related by the CCS algorithm
several spaces that arise naturally in noncommutative geometry (noncommutative
tori, crossed product algebras with the Rokhlin property) and in the theory of
dynamical systems (actions on Cantor sets and associated crossed product
algebras), in the hope that this may allow for the use of techniques from 
noncommutative geometry in the theory of classical and quantum codes. We give
here a quick sketch of what kind of techniques we expect will be applicable in this
context.

\subsection{Spectral triples on fractals and crossed products}

In the recent paper \cite{BeMaRe} the fractals associated to
classical error-correcting codes, as described in \cite{ManMar},
were related to a construction of spectral triples on Cantor sets
of \cite{PeaBell} and to a procedure to obtain crossed product
constructions for such spectral triples. 

Some of the actions of $G_C$ on $\Lambda_C$ described here
are especially suitable for the crossed product construction, as
shown for instance in the recent paper \cite{HawSkaWhiZac}.  This
means that one can regard the crossed product $C(\Lambda_C)\rtimes G_C$
as a a spectral triple (a noncommutative manifold) and apply
to it methods of noncommutative Riemannian geometry. 

There are several interesting constructions of noncommutative geometry,
often in the form of spectral triples, applied to fractal spaces, such as those
obtained in \cite{ChrisIvan}, \cite{ChrisIvanLap}, \cite{GuidoIsola},  \cite{GuidoIsola2}.
It would seem useful to apply some of these techniques to spaces such as
$\Lambda_C$, the subfractals $\Lambda_{C,\ell,\pi}$, or the quotient
$(\Lambda_C\times Q_{\cS,\chi}^*)/\cS$ described above, with the intent
of encoding in the geometry specific information theoretic properties of
both the classical and the quantum code. 

\subsection{Wavelets on fractals and codes}

Representations of Cuntz algebras have been widely used as a way to
construct and analyze wavelets on fractals, see for instance \cite{BraJorg}, 
\cite{DutJorg}, \cite{Jorg1}, \cite{Jorg2}. Some of these constructions, along
with wavelet constructions on fractals obtained in  \cite{Jon} were 
generalized in \cite{MaPa} to the more general case of Cuntz--Krieger algebras. 
Using some of the techniques of \cite{MaPa}, applied to the Cuntz algebras
$\cO_C$ of classical codes, one can similarly obtain a wavelet construction 
on the fractal $\Lambda_C$ associated to the code. This may provide a 
new set of analytic methods to study the decoding procedure in terms of
a wavelet representation. 

\bigskip
\bigskip

{\bf Acknowledgments.} The first author is partially supported by NSF grants
DMS-0901221 and DMS-1007207. The second author was supported for
this project by a Caltech Summer Undergraduate Research Fellowship.

\end{document}